\documentclass[preprint,3p,12pt]{elsarticle}

	\usepackage{amsmath}
	\usepackage{makeidx}
	\usepackage{amsfonts}
	\usepackage[usenames,dvipsnames]{pstricks}
	\usepackage{subfigure}
	\usepackage{epsfig}
	\usepackage{pst-grad} 
	\usepackage{pst-plot} 

\usepackage{color}

\usepackage{latexsym}

    \usepackage[colorlinks=false,linkcolor=black,citecolor=blue,urlcolor=black]{hyperref}
	\usepackage{MnSymbol}

\usepackage{slashed}
\usepackage{color}
\definecolor{orange}{RGB}{255,127,0}

\newcommand{\be}{\begin{eqnarray}}
\newcommand{\ee}{\end{eqnarray}}
\newcommand{\bq}{\begin{equation}}
\newcommand{\eq}{\end{equation}}

\newcommand{\bc}{\begin{center}}
\newcommand{\ec}{\end{center}}

\def\nn{\nonumber}

\begin{document}

\begin{frontmatter}

\title{{\LARGE\sf Dynamics of Majorana fermions in two-dimensions}\\}
\author[umb]{J. A. S\'anchez-Monroy}
\ead{antosan@if.usp.br}
\author[upj]{Abel Bustos}
\ead{abel.alvarez@javerianacali.edu.co}
\address[umb]{Departamento de Ciencias B\'{a}sicas, Universidad Manuela Beltr\'{a}n, Bogot\'{a}, Colombia}
\address[upj]{Departamento de Ciencias Naturales y Matem\'{a}ticas, Pontificia Universidad Javeriana, Cali, Colombia}

\begin{abstract}
A Majorana fermion is the single fermionic particle that is its own antiparticle. Its dynamics is determined by the Majorana equation where the spinor field $(\psi)$ is by definition equal to its charge-conjugate field $(\psi_c)$. Here, we study the dynamics of Majorana fermions in the presence of the most general static external field in $1+1$ dimensions, which is just a scalar potential, by implementing for the first time the methods of supersymmetric quantum mechanics. In particular, for potentials for which shape invariance holds, we show how to obtain analytical solutions. We find that although this equation does have bound states, it does not have stationary states. The approach is illustrated with a linear potential.
\end{abstract}

\end{frontmatter}

\section{Introduction}
The Dirac equation is the relativistic generalization of the Schr\"{o}dinger equation \cite{dirac1928quantum}. This equation successfully merges quantum mechanics with special relativity for spin-$1/2$ particles and predicts the existence of the antiparticle. In four dimensions the \textit{Dirac spinor} ($\psi_D$) is an object with four degrees of freedom corresponding to a particle and an anti-particle with spin up ($\uparrow$) and down ($\downarrow$). For an electrically neutral particle, however, it is possible that it be its own antiparticles; in which case, the spinor has only half the degrees of freedom compared to the Dirac fields. This kind of particle is referred to as \textit{Majorana fermion} and was proposed by Ettore Majorana in 1937 \cite{majorana1937teoria}. 
\par
To avoid confusion with the nomenclature used in the literature, let us clarify that Lorentz invariance allows, in addition to the Dirac equation $i\hbar\slashed{\partial}\psi= mc\psi$, the equation known as the Majorana equation $i\hbar\slashed{\partial}\psi= mc\psi_c$ \cite{majorana1937teoria,zee2010quantum}. Here $\psi_c$ stands for charge conjugation of the spinor $\psi$. When the additional condition $\psi_c=\psi$ is imposed in the Majorana equation, the resulting spinor is known as a Majorana spinor (or Majorana fermion). Indeed, Majorana fermions can also be viewed as real solutions of the Dirac equation when the so-called Majorana representation is chosen \cite{wilczek2009majorana}. To distinguish them from a field that obeys the Majorana equation without the condition $\psi_c=\psi$, the latter have been called \textit{Majoranons} \cite{Changsuk2013}. 
\par
After the discovery of neutrino masses and mixing, the Majorana fermions have attracted renewed interest since in the Standard Model of particles it is not known whether neutrinos are their own antiparticles. On the other hand, the notion of Majorana fermions has a realization in topological superconductors \cite{Alice2012,Beenakker2013}, and it has potential applications in quantum computation \cite{Kitaev2001,Tewar2013}. Besides, in recent years, there has been a growing interest in simulations of relativistic quantum systems such as the Dirac \cite{gerritsma2010} and Majorana equations \cite{Casanova2011a,Changsuk2012a,Liu2014,Lara2014a,Robert2015a,zhang2015a} in $1+1$ dimensions. Interestingly, some peculiar effects predicted by the Dirac equation, such as Klein tunneling and ``Zitterbewegung'' (a trembling motion caused by the interference between positive and negative energy states \cite{schrodinger1930}), unobservable in experiments of high energy physics could be observable in
quantum simulations \cite{gerritsma2010,Casanova2010a,Gerritsma2011a}. Furthermore, quantum simulations of the Majorana equation have opened the possibility for implementing unphysical operations, such as complex conjugation, charge conjugation and time reversal \cite{Casanova2011a,zhang2015a,Robert2015a}.
\par
While the analytical solutions of the relativistic equations of Dirac and Klein-Gordon have been widely investigated in $d$ dimensions \cite{bagrov1990,thaller1992,cooper1995,Sanchez2014}, the solution of the Majorana equation remains essentially unexplored \cite{2012Lamata,2017Hashimia,Casanova2012a}. In \cite{2012Lamata}, the authors studied the nonrelativistic limit of the Majorana equation in $1+1$ dimensions and show that in the non-relativistic limit the dynamics is a superposition of the positive and negative ``energy branches''; in contrast to what happens with the non-relativistic limit of the Dirac equation where the positive and negative energy branches do not mix. In general, Majorana fermions \cite{2017Hashimia} and Majoranons \cite{2012Lamata} have not stationary energy eigenstates. In \cite{2017Hashimia} it was studied a Majorana spinor confined to one-dimensional interval with the most general perfectly reflecting boundary conditions, which are described by only two discrete types of wall boundary conditions rather than a continuous $1$-parameter family of parameters as in the case of a Dirac fermion. Recently it was realized that, for a certain class of potentials, the dynamics of a Majoranons in $1+1$ dimensions can be mapped to the solution of the free Majorana equation \cite{Casanova2012a}.
\par
Motivated by these counterintuitive features and by the few known analytical solutions, this paper aims to provide a method to obtain exact solutions of Majorana fermions in the presence of the most general static external field in $1+1$ dimensions. To this end, we will show that this problem is equivalent to solving a problem in supersymmetric quantum mechanics (SUSY-QM) \cite{Witten1981}. The techniques of SUSY-QM are basically equivalent to the so-called factorization method \cite{Infeld1951}. This elegant method allows finding a pair of isospectral Hamiltonians. Moreover, if this pair is invariant under a discrete reparametrization (shape invariance), the Hamiltonians can be solved exactly \cite{gendenshtein1983}. Although these techniques were initially developed in the context of non-relativistic quantum mechanics, they have been widely used to solve the Klein-Gordon \cite{CHEN2006,JANA2007,Setare2009,Dong2011} and Dirac \cite{Sanchez2014,Cooper1981a,Debergh2002,ZARRINKAMAR2010} equations. 
\par
This paper is organized as follows. In Sec. \ref{Sec2} we introduce the most general external potential in the two-dimensional Majorana equation. In Sec. \ref{Sec3} we describe the methods of supersymmetric quantum mechanics and implement them in the Majorana equation. In Sec. \ref{Sec4}, we present the explicit solution of the Majorana equation in the presence of a linear potential, where the absence of stationary states for Majorana fermions is emphasized. Finally, Sec. \ref{Sec5} contains our conclusions.
\section{Majorana fermions in $(1+1)$ dimensions} \label{Sec2}
In $(1+1)$ dimensions the free Dirac equation is ($\mu=0,1$)
\begin{equation}\label{ED}
\left[i\gamma^{\mu}\hbar\partial_{\mu}-mc\right]\Psi_D(x,t)=0,
\end{equation}
where the $\gamma$-matrices are the generators of the two-dimensional Clifford algebra
given by
\begin{equation}
\{\gamma^{\mu},\gamma^{\nu}\}=2\eta^{\mu \nu}, \ \ \ \ \text{with} \ \ \ \ \eta^{\mu \nu}=diag(+,-).
\end{equation}
Here the $\gamma$-matrices are $2 \times 2$ dimensional matrices
for the irreducible representation. The spinor $\Psi_D$ is a two component complex valued function which has two degrees of freedom associated with particles and antiparticles.  
\par
The Majorana modification of the Dirac equation consists in constructing an equation in which the fermions particles are their own antiparticles\footnote{It turns out that there is no such thing as spin in two dimensions, therefore strictly speaking the Dirac equation in $(1+1)$ dimensions does not describe half-integer spin particles.} \cite{majorana1937teoria}. Consequently, the Majorana spinor ($\Psi_M$) will have only half the degrees of freedom compared to the Dirac fields. In a particular representation of the matrices, known as the Majorana representation, this statement translates into the requirement of reality, $\Psi_M^*(x,t)=\Psi_M(x,t)$ \cite{wilczek2009majorana,2017Hashimia}.
Therefore, the Majorana spinor is a two component real valued function. 
\par
The Majorana representation is chosen as
\begin{eqnarray}\label{MajoranaBasis}
\gamma^{0}=-\sigma^2=\left( \begin{array}{cc}
0 & i \\
-i & 0 \\
\end{array} \right), \ \ \ \ \  \gamma^{1}=i\sigma^3=\left( \begin{array}{cc}
i & 0 \\
0 & -i \\
\end{array} \right), \ \ \ \ \
\gamma^{5}=i\gamma^{0}\gamma^{1}=i \sigma^1=\left( \begin{array}{cc}
0 & i \\
i & 0 \\
\end{array} \right).
\end{eqnarray}
The (1+1)-dimensional Dirac equation in the presence of the most general static external fields is
\begin{eqnarray}
\left[i\gamma^{\mu}(\hbar\partial_{\mu}+i\frac{e}{c}A_{\mu}(x))-\frac{1}{c}P(x)
\gamma^{5}-\frac{1}{c}V(x)\mathbf{1}-mc\mathbf{1}\right]\Psi_D(x,t)=0.
\end{eqnarray}
In the last expression, $\mathbf{1}$ stands for the $2\times 2$ identity
matrix. Because there are four linearly independent $2\times 2$ matrices, $\mathbf{1}$, $\gamma^0$, $\gamma^1$ and $\gamma^5$, there are four external potentials. The external potentials can be identified as follows: the scalar potential $V(x)$, the pseudoscalar potential $P(x)$ and the gauge potential $A_{\mu}(x)=\{A_t(x),A_x(x)\}$. 
\par
The Majorana equation in the presence of a external potential can be written as 
\begin{equation}
\left[i\gamma^{\mu}\hbar\partial_{\mu}-mc\mathbf{1}-\frac{1}{c}\mathcal{V}\right]\Psi_M(x,t)=0.
\end{equation}
One might think that $\mathcal{V}$ be a linear combination of the four $2 \times 2$ matrices
\begin{equation}
\gamma^0\mathcal{V}=\mathbf{1} f_1(x)+\gamma^0 f_2(x)-i\gamma^1 f_3(x)-i\gamma^5 f_4(x).
\end{equation}
However, as we will see, this is not the case. Let us rewrite the Majorana equation  in the form of
a Schr\"{o}dinger equation
\begin{eqnarray} \nn
i\hbar\frac{\partial}{\partial 
t}\Psi_M(x,t)&=&\mathcal{H}\Psi_M(x,t)\\\label{SchrMajorana}
&=&\left(c\alpha \hat{p}+ \beta
mc^2+\gamma^0\mathcal{V}\right)\Psi_M(x,t), 
\end{eqnarray}
where $\mathcal{H}$ is the (Dirac) Hamiltonian operator, $\hat{p}$ is the momentum operator, $\beta=\gamma^{0}$ and $\alpha=\gamma^{0}\gamma^{1}$. Since $\mathcal{H}$ must be Hermitian, the functions $f_i(x)$  must be real valued functions. Notably, the requirement of reality ($\Psi_M^*=\Psi_M$) imposes that $f_1(x)=f_3(x)=f_4(x)=0$, as can be easily verified taking the complex conjugate of Eq. (\ref{SchrMajorana}). Since the potentials $f_2$ can be identified as a scalar potential, $f_3$ as a pseudoscalar one and $(f_1,f_4)$ as a gauge potential,  
the most general external potential in the Majorana equation is just a scalar potential (in the following $f_2(x)=\phi(x)$),
\begin{equation}
\left[i\gamma^{\mu}\hbar\partial_{\mu}-mc-\frac{1}{c}\phi(x)\right]\Psi_M(x,t)=0.
\end{equation}
While the condition $f_1=f_4=0$ is obvious since the Majorana fermions have to be neutral, the condition $f_3=0$ is unexpected because it prevents Majorana fermions coupling to a pseudoscalar potential. 
\par
The previous result has an important consequence. The presence of an electric potential in the Dirac equation can lead to the existence of the so-called Klein paradox (instability of Dirac sea) \cite{Sanchez2014}. It turns out that the Klein tunneling causes that potentials which are confining in the non-relativistic limit to be nonconfining in the relativistic limit. From the point of view of quantum field theory, the Klein paradox is explained by the creation of a particle-antiparticle pair by an external potential. Since Majorana fermions are its own antiparticles and it is not possible for them to interact with an electric field, there is no Klein tunneling.
\section{Supersymmetric quantum mechanics for Majorana fermions}\label{Sec3}
Now, we present the method to solve the Majorana equation. In the Majorana representation (\ref{MajoranaBasis}), we can rewrite Eq. (\ref{SchrMajorana}) as 
\begin{eqnarray}\label{Majotwocomp}
\hbar \partial_t \psi_1 &=& \left[-c\hbar  \partial_x +mc^2+\phi(x)\right]\psi_2,\\
-\hbar \partial_t \psi_2 &=& \left[+c\hbar  \partial_x +mc^2+\phi(x)\right]\psi_1,
\end{eqnarray}
where $\Psi_M=(\psi_1,\psi_2)^T$. The above equations can be written in terms of the first-order differential operators $A$ and $A^{\dag}$, which are known as ladder operators, 
\begin{eqnarray}
\hbar \partial_t \psi_1= \hat{A}^{\dag}\psi_2, \label{APAP1}\\
-\hbar \partial_t \psi_2= \hat{A}\psi_1,\label{APAP2}
\end{eqnarray}
where the ladder operators are defined as
\begin{eqnarray}
\hat{A}&=&c\hbar  \partial_x +mc^2+\phi(x),\label{esps1}\\
\hat{A}^{\dag}&=&-c\hbar  \partial_x +mc^2+\phi(x).\label{esps2}
\end{eqnarray}
The equation system (\ref{APAP1}) and (\ref{APAP2}) can be decoupled into the two following second-order differential equations
\begin{eqnarray}
&&-\hbar^2\partial_{t}^2\psi_1=\hat{A}^{\dag}\hat{A}\psi_1=\hat{H}_{-}\psi_1, \label{mkge1}\\
&&-\hbar^2\partial_{t}^2\psi_2=\hat{A}\hat{A}^{\dag}\psi_2=\hat{H}_{+}\psi_2. \label{mkge2}
\end{eqnarray}
The effective supersymmetric partners $\hat{H}_{-}$ and $\hat{H}_{+}$ are given by
\begin{eqnarray}\label{effecsuperhamils}
\hat{H}_{\pm}=-(c\hbar)^2\frac{d^2}{dx^2}+V_{\pm},
\end{eqnarray}
with the so-called supersymmetric partner potentials $V_{\pm}=(mc^2+\phi)^2\pm c\hbar \frac{d \phi}{dx}$.
Since $\hat{H}_{\pm}$ are time-independent and $\psi_{1,2}$ are real valued functions, using a separation ansatz $\psi_{1,2}(t,x)=f(t)_{\mp}\varphi_{\mp}(x)$, we find that the solutions are of the form
\begin{eqnarray}\label{psi1sin}
\psi_{1}(t,x)&=&\varphi^{-}(x)\sin \left[\frac{\mathcal{E} t}{\hbar}+\delta_{-}\right],\\\label{psi2cos}
\psi_{2}(t,x)&=&\varphi^{+}(x)\cos \left[\frac{\mathcal{E} t}{\hbar}+\delta_{+}\right],
\end{eqnarray}
with $\delta_{\pm}$ free parameters and $\varphi^i(x)$ ($i=\pm$) the solutions of the equation
\begin{eqnarray}\label{espectrumh2+-}
\mathcal{E}^2 \varphi^{\pm}(x)=\hat{H}_{\pm}\varphi^{\pm}(x).
\end{eqnarray}
Employing Eqs. (\ref{APAP1}) and (\ref{APAP2}), we find that $\delta_{-}=\delta_{+}=\delta$ and that for each eigenstate with $\mathcal{E} \neq 0$ there is a one-to-one mapping between the energy eigenstates $\varphi^{\pm}$:
\begin{eqnarray}\label{Ladderrelation}
\hat{A}^{\dag} \varphi^+=\frac{1}{\mathcal{E}}\varphi^-,\ \ \ \ \ \hat{A} \varphi^-=\frac{1}{\mathcal{E}}\varphi^+.
\end{eqnarray}
Consequently, solving the Majorana equation in the presence of any static external field is equivalent to solving a problem of supersymmetric quantum mechanics. 
\par
When one of the Hamiltonians has a zero energy eigenstate ($\mathcal{E}=0$), SUSY is called unbroken. Otherwise, SUSY is called broken \cite{cooper1995}. In the unbroken case, one of the Hamiltonians has an additional eigenstate at zero energy that does not appear in its partner Hamiltonian. From Eq. (\ref{Ladderrelation}) the zero energy eigenstate of $\hat{H}_+(\hat{H}_-)$ can be determined imposing that it is annihilated by the operator $\hat{A}^{\dag}(\hat{A})$.
\par
If $\hat{A}\varphi^{-}_{0}=0$ and $\varphi^{-}_{0}$ is normalizable, one can show that \cite{cooper1995,2011supersymmetric}
\begin{eqnarray}
\mathcal{E}^{+}_n&=&\mathcal{E}^{-}_{n+1}, \ \ \ \mathcal{E}^{-}_0=0, \\ \label{Avarphi+re}
\varphi^{+}_{n}&=& [\mathcal{E}^{-}_{n+1}]^{-1}\hat{A}\varphi^{-}_{n+1},\ n=0,\ 1,\ 2\ \ldots, \\
\varphi^{-}_{n+1}&=&[\mathcal{E}^{+}_{n}]^{-1}\hat{A}^{\dag} \varphi^{+}_{n},\ n=1,\ 2\ \ldots.
\end{eqnarray}
If instead $\hat{A}^{\dag}\varphi_{+}^{0}=0$ and $\varphi_{+}^{0}$ is normalizable, one obtain the following relations
\begin{eqnarray}
\mathcal{E}^{-}_n&=&\mathcal{E}^{+}_{n+1}, \ \ \ \mathcal{E}^{+}_0=0, \\
\varphi^{-}_{n}&=&[\mathcal{E}^{+}_{n+1}]^{-1}\hat{A}^{\dag}\varphi^{+}_{n+1},
\ n=0,\ 1,\ 2\ \ldots, \\
\varphi^{+}_{n+1}&=&[\mathcal{E}^{-}_{n}]^{-1}\hat{A}\varphi^{-}_{n},\ n=1,\ 2\ \ldots.
\end{eqnarray}
We note that in general the solutions of the Majorana equation do not yield stationary energy eigenstates, as can be checked by computing the probability density $\rho_n(t,x)=[\Psi_n(t,y)]^T\Psi_n(t,y)$, the ground state is the exception. In fact, the latter is always true for unbroken SUSY, because in this case for ground state $\mathcal{E}=0$.
\subsection{Shape invariance}
Although the supersymmetry tells us that the partner Hamiltonians are
(almost) isospectral, it does not tell us how to determine the spectrum. 
However, when the supersymmetric Hamiltonians depends on a real parameter in
such a way that these are invariant under a discrete reparametrization (shape invariance), it is possible to obtain the spectrum algebraically \cite{gendenshtein1983}. 
\par
If $\hat{H}_{-}$ has a zero eigenvalue, the shape invariance condition is given by
\begin{equation}
\hat{H}_{+}(a_1,x)=\hat{H}_{-}(a_2,x)+R(a_1),
\end{equation}
i.e. the two supersymmetric partner Hamiltonians have the same form, but they differ in a constant ($R$) and in dependence on $a_i$. Here, the parameter $a_2$ is a function of the parameter $a_1$, $a_2=f(a_1)$. One can construct a sequence of Hamiltonians by applying successive reparametrizations $a_s=f^{s-1}(a_1)$, and using the condition of shape invariance, thus, we have
\begin{equation}
\hat{H}_{+}(a_{s-1},x)=\hat{H}_{-}(a_{s},x)+R(a_{s-1}).
\end{equation}
Thus, the spectrum of (\ref{espectrumh2+-}) can be easily found \cite{2011supersymmetric}
\begin{eqnarray}\label{EigenE-Ma}
\mathcal{E}_{n}^-(a_1)=\pm\sqrt{\sum_{k=1}^{n}R(a_k)}, \ \ \ \ \mathcal{E}_{0}^-(a_1)=0.
\end{eqnarray}
It should be noted that although $\mathcal{E}$ represents the eigenvalues of $H_{\pm}$, it is not an eigenvalue of the Majorana equation, in which there are not energy eigenstates associated with a
unique positive or negative energy \cite{2012Lamata,2017Hashimia}. Furthermore, the two signs of Eq. (\ref{EigenE-Ma}) lead us to the same state, as we will discuss later.

\section{Example: Linear potential}\label{Sec4}
Let us illustrate the method developed above with a simple example, a scalar linear potential 
\begin{equation}
\phi(x)=kx,
\end{equation}
with $k$ a constant. For this system, the supersymmetric partners Hamiltonians, Eq. (\ref{effecsuperhamils}), are given by
\begin{eqnarray}
H_{-}&=&-(c\hbar)^2\frac{d^2}{dx^2}+m^2c^4+k ^2 x^2+2 m c^2 k x-c\hbar
k,\\ 
H_{+}&=&-(c\hbar)^2\frac{d^2}{dx^2}+m^2c^4+k ^2 x^2+2 m c^2 k x+c\hbar k.
\end{eqnarray}
If $a_s=k$ and $R(a_s)=2c\hbar k$, the shape invariant condition is fulfilled. Assuming that $H_{-}$ have an eigenstate at zero energy, Eq. (\ref{EigenE-Ma}) yields
\begin{equation}
\mathcal{E}_{n}^-=\pm\left(\sum^{n}_{s=1}2c\hbar k\right)^{\frac{1}{2}}=\pm \sqrt{2c\hbar k n}, \ \ \ \ \ \mathcal{E}_{0}^-=0.
\end{equation}
In this case $\hat{A}\varphi^{-}_{0}=0$ and hence the ground state given by 
\begin{eqnarray}
\varphi^{-}_{0}(y)=\left(\frac{w}{\pi}\right)^{\frac{1}{4}}\exp\left(-\frac{w}{2}y^2\right),
\end{eqnarray}
with $w=k/(c\hbar)$ and where we made the change of variable $y=x+mc^2/k$. Notice that normalizability of $\varphi^{-}_{0}(y)$ requires that $k>0$. We can now obtain the excited states by successive applications of $\hat{A}^{\dag}$ \cite{cooper1995}
\begin{eqnarray}\nn
\varphi^{-}_{n}(y)&\sim&(\hat{A}^{\dag})^n\varphi^{-}_{0}(y)\\\label{phi-nlinear}
&=&\frac{1}{2^{\frac{n}{2}}(n!)^{\frac{1}{2}}}\left(\frac{w}{\pi}\right)^{\frac{1}{4}}\exp\left(-\frac{w}{2}y^2\right)H_{n}[w^{\frac{1}{2}}y],
\end{eqnarray}
where $H_{n}$  are the Hermite polynomials. By inserting Eq. (\ref{phi-nlinear}) into (\ref{Avarphi+re}), we can obtain the eigenstates of $H_{+}$
\begin{eqnarray}\nn
\varphi^{+}_{n}(y)&=&\frac{1}{2^{\frac{n-1}{2}}((n-1)!)^{\frac{1}{2}}}\left(\frac{w}{\pi}\right)^{\frac{1}{4}}\exp\left(-\frac{w}{2}y^2\right)H_{n-1}[w^{\frac{1}{2}}y].
\end{eqnarray}
Employing Eq. (\ref{psi1sin}) and (\ref{psi2cos}), we finally get 
\begin{eqnarray}\label{SolutionMajoranaLineal}
\Psi_n(t,y)=\frac{w^{\frac{1}{4}}e^{-\frac{w}{2}y^2}}{2^{\frac{n}{2}}(n!)^{\frac{1}{2}}\pi^{\frac{1}{4}}}\left(\begin{array}{c}
H_{n}[w^{\frac{1}{2}}y] \sin \left[c(2 w n)^{\frac{1}{2}}t+\delta\right] \\
\sqrt{2 n} H_{n-1}[w^{\frac{1}{2}}y] \cos \left[c(2 w n)^{\frac{1}{2}}t+\delta\right] \\
\end{array}\right),
\end{eqnarray}
for $n=1,2,\dots$ and 
\begin{eqnarray}
\Psi_0(t,y)=\frac{w^{\frac{1}{4}}e^{-\frac{w}{2}y^2}}{\pi^{\frac{1}{4}}}\left(\begin{array}{c}
1 \\
0 \\
\end{array}\right),
\end{eqnarray}
\begin{figure*}
\center
\includegraphics[width=16.6cm]{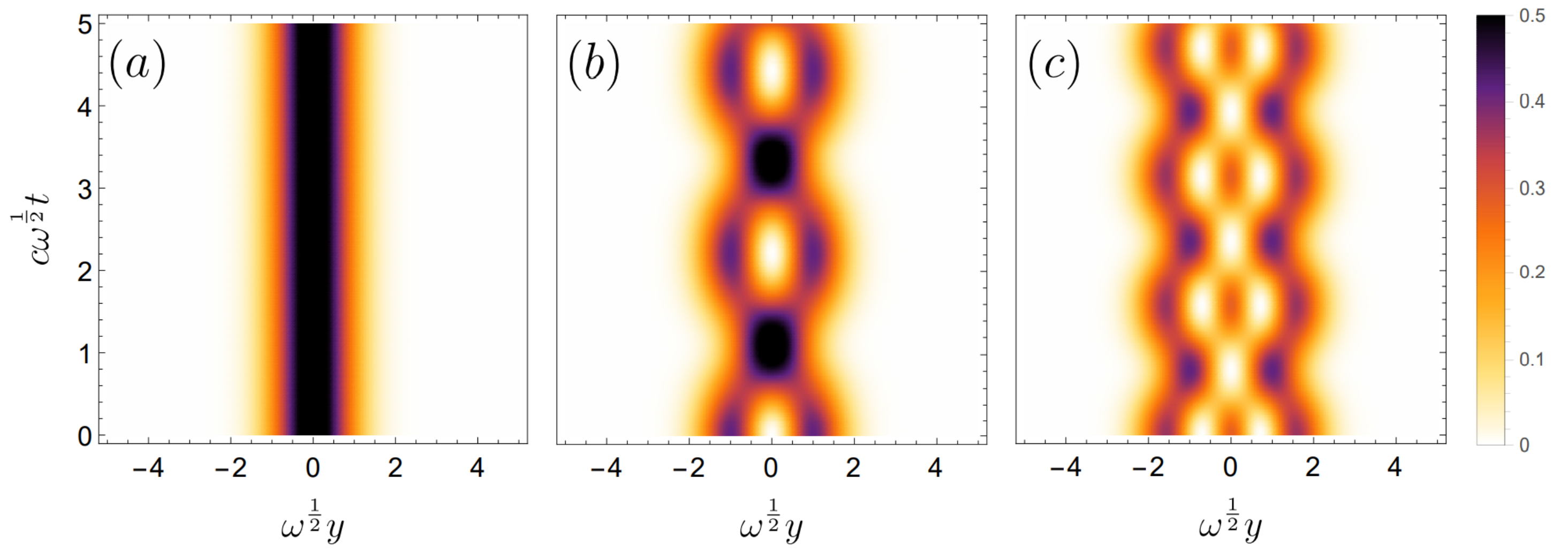}
\caption{Evolution of the probability density under a linear potential for $\delta=\pi/2$ and $n=0$ (panel a), $n=1$ (panel b), $n=2$ (panel c). The probability density is given in units of $\omega^{-\frac{1}{2}}$.}
\label{DensityPlot}
\end{figure*}
for $n=0$. Here, the Majorana spinor is normalized as $\int \Psi^T\Psi dx=1$. Since Eq. (\ref{SolutionMajoranaLineal}) was written for positive values of $\mathcal{E}$, we also might have considered the negative values of $\mathcal{E}$. However, using Eq. (\ref{Ladderrelation}), one can prove that the solutions are
\begin{eqnarray}
\Psi_n^{-}(t,y)=\frac{w^{\frac{1}{4}}e^{-\frac{w}{2}y^2}}{2^{\frac{n}{2}}(n!)^{\frac{1}{2}}\pi^{\frac{1}{4}}}\left(\begin{array}{c}
H_{n}[w^{\frac{1}{2}}y] \sin \left[-c(2 w n)^{\frac{1}{2}}t+\delta_{-}\right] \\
-\sqrt{2 n} H_{n-1}[w^{\frac{1}{2}}y] \cos \left[-c(2 w n)^{\frac{1}{2}}t+\delta_{-}\right] \\
\end{array}\right),
\end{eqnarray}
which are not linearly independent of (\ref{SolutionMajoranaLineal}) putting $\delta=-\delta_{-}$. Consequently, there is only one ``family'' of solutions. 
\par
In contrast to the solutions of the Dirac equations, the solutions of the Majorana equations depend on a free parameter $\delta$. To understand the origin of this parameter, let us observe that for the Dirac equation for a stationary energy eigenstates is of the form $\Psi(t,x) =e^{-iEt/\hbar}u(x)$, with $u(x)$ a spinor. Since one can always shift the time coordinate $t\rightarrow t-t_0$, such that the probability density remains unchanged, the solutions of the Dirac equation for a stationary state can be rewritten as $\Psi(t,x) =e^{-iE(t-t_0)/\hbar}u(x)$, without changing any physical observable. Of course, this is nothing more than the global $U(1)$ symmetry of the Dirac equation. Because the Majorana equation is not invariant under multiplication of $\Psi$ by an arbitrary $U(1)$ element \cite{2017Hashimia}\footnote{The Majorana equation only is invariant under a discrete group $\mathbb{Z}_2$ i.e. invariant under the multiplication by $1$ and $-1$.}, a shift in the time coordinate does not leave invariant the probability density. This, together with the fact that there are not stationary states, leads to that there is freedom when we choose our initial condition. This freedom is incorporated into the $\delta$ parameter. 
\par
When $k<0$ the normalizable solution is $\varphi^{+}_{0}(y)$ and $H_{+}$ have an eigenstate at zero energy. The solutions, in this case, can be obtained by exchanging rows at (\ref{SolutionMajoranaLineal}) and making $w\rightarrow -w>0$.
\par
In Fig. \ref{DensityPlot}, we show the evolution of the probability density for $n=0$, 1, and 2 ($\delta=\pi/2$) in the panels (a), (b) and (c), respectively. The evolution is periodic for $n>0$, with period equal to $2^{\frac{1}{2}} \pi c^{-1}(\omega n)^{-\frac{1}{2}}$. Although there are no stationary states, the solutions of the Majorana equation in a linear potential remain localized in a region of space, such that the probability density tends to zero as $y\rightarrow \pm \infty$, in other words, the solutions are bound states.
\par
Recently, the Dirac equation in the presence of a linear scalar potential was solved analytically in the Majorana representation \cite{Ribeiro2017}. Because the requirement of reality was not implemented, the solutions found there do not correspond to Majorana fermions, but to Dirac fermions (compare, for example, with Ref. \cite{Sanchez2014} where the Dirac problem was studied).
\section{Conclusions}\label{Sec5}
In this paper, we studied the dynamics of Majorana fermions, i.e., particles that obey the Majorana equation and the Majorana condition $\psi_c=\psi$, in two-dimensions. It was found that the Majorana condition implies that the most general external potential with which Majorana fermions can interact is simply a scalar potential. We have shown how SUSY-QM can be used to solve the Majorana equation in the presence of a static scalar potential. Although for unbroken SUSY the ground state is a stationary state, in general, any other state is not stationary. Furthermore, we used the shape invariance property to find analytical solutions algebraically, highlighting the differences with the solutions for Dirac fermions. 
We exemplify our approach with a linear potential. However, it can be used to obtain analytical solutions for a large family of potentials such as P\"{o}schl-Teller, Rosen-Morse and Scarf potentials, which have been studied within the supersymmetric quantum mechanics.
\section*{Acknowledgements}
The authors are grateful to Y. Perez for useful comments. 

\end{document}